# Interdisciplinary patterns of a university
# Investigating collaboration using co-publication network analysis

Uwe Obermeier, Hannes Brauckmann
Innovation Research Unit
University College Dublin
Corresponding author: uwe.obermeier@ucd.ie

**Abstract:**

We investigate collaborative and interdisciplinary research features of University College Dublin, using methods from social network analysis to analyze and visualize (co-)publications covered by the Web of Science from 1998 through 2007. We account for the extent of interdisciplinarity in collaborations, distinguishing collaborations between schools within one college ("small interdisciplinarity") from collaborations between schools in different colleges ("big interdisciplinarity"). Based on the interdisciplinary nature, we compare the types of collaboration to a model of random matching across units, observing several marked differences. During the period of consideration, collaborations within UC Dublin nearly doubled, almost entirely due to the increasing level of intra-school collaborations.

# The "network mode" of knowledge production

## 1. Introduction

So far, various theoretical frameworks have been used to describe recent changes in academic research. Though these frameworks differ in many respects, they all point to a more collaborative "network mode" of knowledge production. Concepts such as "Mode 2" (Gibbons et al., 1994), "Academic Capitalism" (Slaughter and Leslie, 1997), "Post-Academic Science" (Ziman, 2000), or "Triple Helix" (Etzkowitz and Leydesdorff, 2000) not only refer to external collaborations of universities with industry, government and other actors, but report changed practices inside academia.

In universities, the isolated researcher in the ivory tower has been widely replaced by teams in collaborative research projects (Wuchty et al, 2007). Crossovers and co-operations between different scientific disciplines, different organizational units, and external actors seem to be a common and increasing phenomenon of academic reality (Guimerà et al, 2005). Since collaborative research has become the dominant and most promising way to produce high-quality output (Jones et al, 2008), collaboration structures are also a target for research and management design (Bozeman, Lee 2005). Collaborative research projects, co-authored publications, or multidisciplinary excellence networks in universities point to the peer network mode of today's knowledge production.

Interdisciplinarity is considered to be essential for the advancement of science. As not only universities (Bordons et al, 1999) foster interdisciplinary research, so too do funding agencies at national and European levels, so knowledge relating to existing interdisciplinary links will become increasingly interesting for those involved. The reasons for this are twofold: on the one hand, these kinds of analyses can provide a map beyond the rhetoric of interdisciplinarity and provide measurements of the existing links. On the other hand, this information becomes even more valuable when an organization applies for funding for further interdisciplinary research. It is helpful to know and to demonstrate that interdisciplinary collaboration already exists, and these links should be nurtured rather than built from scratch. This knowledge of past collaboration is especially important given that empirical findings about group collaboration show that people who already have written a paper previously together are much more likely to succeed in future collaborations. These researchers have already paid the start-up costs of getting to know each other's languages, approaches and methodologies (Cummings and Kiesler 2008).

The study is set up as an academic research project but is additionally intended to contribute to the self-monitoring mechanisms of this university as well. We assume that demand from academic management for consultancy will grow due to rising legitimatory needs.

## 2. Data

Data from our study reflects the publication output of University College Dublin's (UCD) permanent academic staff published during the period 1998-2007. These include scientific articles published in journals, proceedings and serials processed for the Web of Science (WoS) versions of Science Citation Index and associated indices: the Science Citation Index (SCI), Social Science Citation Index (SSCI), and the Arts and Humanities Index (A&HCI). The data was provided from a internal database of university management where the output of all academic staff of the university is recorded.

In UCD, all members of academic staff belong to schools which are organised in a college structure. There are five different colleges and thematically close schools and institutes (we treat institutes as schools) are related to these colleges: *College of Life Sciences* with eight related schools and five institutes, *College of Business and Law* with only two schools and one institute, *College of Arts & Celtic Studies* with eight related schools and two institutes, *College of Human Sciences* with eight associated schools and one institute and the *College of Engineering, Mathematical & Physical Sciences* with seven related schools.

## 3. Method and Results

The combination of co-authorship and Social Network Analysis (SNA) is often used for mapping, analysing, and evaluating scientific research activity (Olmeda-Gómez et al 2008, Barabasi et al 2002, Yousefi-Nooraie et al 2008).

Different approaches to measure interdisciplinarity exist. Measures like the (co-) classifications of journals have widely been used to produce maps of specific research areas (Katz and Hicks 1995). Approaches to study interdisciplinarity between different disciplines and subfields also have been used (Morillo, Bordons, Gomez 2003). One should keep in mind that subject categories refer to journals and their classifications, however the co-authorship between different organizational units of a university, is not shown. But due to the fact that we were especially interested in the collaborative links within this particular university, we propose the combination of social network analysis, bibliometrics and organizational units as an approach to measure the (interdisciplinarity) collaboration within a particular university.

Our analysis is divided into two parts, one dealing with collaboration numbers and rates of the UCD colleges and their evolution over time and the other investigating the co-publication activity of specific school- and college-pairings. First we describe the method of part one and present the results for the colleges. Subsequently the method of the second part is described followed by the associated results. Finally we present the evolution of collaboration activity within UCD computed according to method one.

**Methods I: Collaboration per college and evolution of co-publication**

For the purposes of our study, we measure interdisciplinarity as co-publication activity of UCD authors belonging to different Colleges and Schools. Following the approach used from Morillo et al 2003, and their classification of journals as "big" interdisciplinarity and "small" interdisciplinarity we are considering co-publications between schools of the same college as "small" interdisciplinarity and co-publications between schools of different colleges as "big" interdisciplinarity. The schools are ordered thematically to the college structure.

In order to study both the distribution over the colleges and the evolution of the co-publication activity within UCD the following analysis was carried out using a self-made Python script:
First the database was transformed to a co-authorship network by indentifying all UCD authors of each publication and taking every author-pairing as a co-authorship link afterwards. The co-authorship links are cumulated over all publications. Additionally the following numbers and rates of publications (per year / that include one author from the college) were computed for each year and college respectively:
1) The number of all publications which can be used to infer rates of co-publication.
2) The collaboration rate i.e. the percentage of publications written by at least two authors within UCD which stands for the total amount of co-publication activity.
3) The intra-school collaboration rate i.e. the percentage of publications written by at least two authors affiliated to only one school. This gives a measure of intra-school co-publication activity.
4) The inter-school collaboration rate i.e. the percentage of publications written by at least two authors from different schools of the same college. We consider this as a measure of "small" interdisciplinary collaboration activity.
5) The inter-college collaboration rate i.e. the percentage of publications written by at least two authors from different colleges. Finally this gives a measure of "big" interdisciplinary collaboration activity.

It is important to note that the collaboration rates 3) 4) and 5) are independent of each other and add up to the total collaboration rate 2).

**Results I: Collaboration analysis per college**

In Figure 1 we only focus on researchers with publications co-authored within UCD. Every node represents a researcher with at least one co-authored publication (Table 1, row 2), a link between two nodes represents at least one co-authored publication. This includes co-authorship within the same school as well as co-authorship between different schools of the same college (intra-college links) and co-authorship between different schools of different colleges (inter-college links). We can observe two highly connected clusters from the *College of Life Sciences.* The highest number of researchers with co-authorship outside their own college we can observe within the *College of Engineering, Mathematical & Physical Sciences* with 23, and *College of Life Sciences* with 42 researchers. For the *College of Business and Law* the number of researchers co-publishing outside their own college is small 4, but the percentage 40 % is high (see Table 1, row 3).

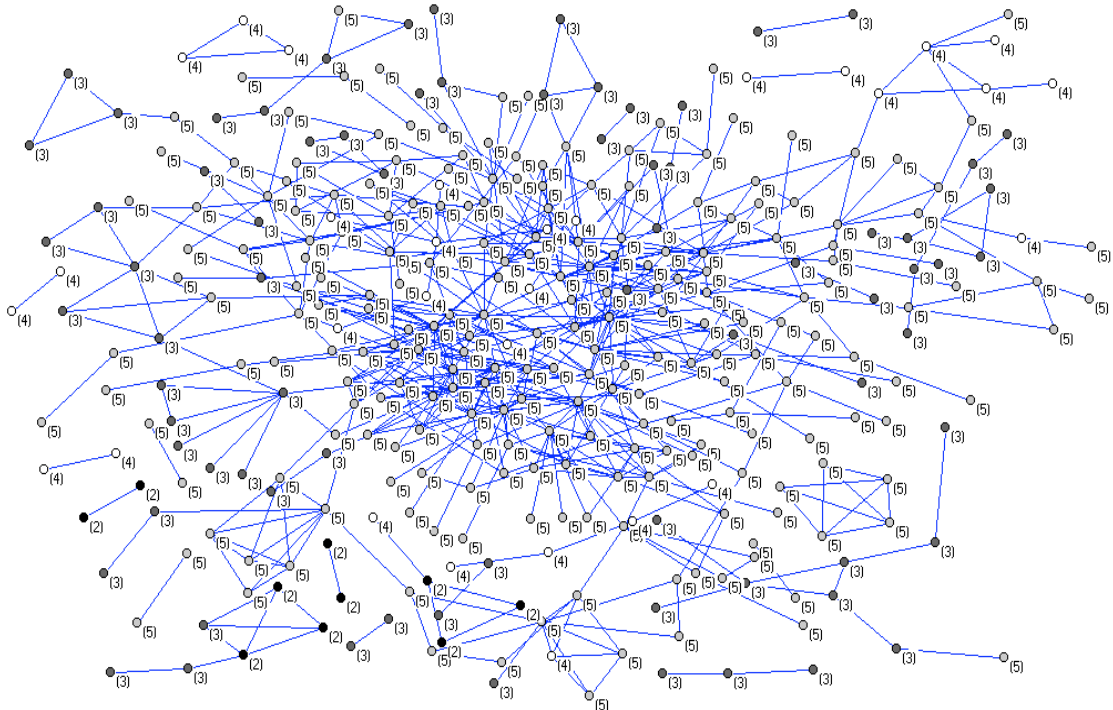

Fig 1: Co-authorship network of the 341 researchers with co-publications within and between colleges 1998-2007. The largest component contains 234 authors. Different levels of gray were used to distinguish the colleges. Nodes (2) College of Business & Law, Nodes (3) College of Engineering, Mathematical & Physical Sciences, Nodes (4) College of Human Sciences, Nodes (5) College of Life Sciences. A database covering 641 researchers and 7836 publications was analyzed. Data Source: UCD Research, Software used: Pajek

| College | Arts (1) | Business (2) | Engineering (3) | Human (4) | Life Sciences (5) |
|---|---|---|---|---|---|
| Authors represented in WoS | 61 | 44 | 140 | 99 | 297 |
| Authors with Co-authorship in UCD | / | 10 (23%) | 75 (54%) | 30 (30%) | 226 (76%) |
| Co-authorship outside own college | / | 4 (40%) | 23 (31%) | 5 (17%) | 42 (19%) |

Tab 1: Author numbers per college. The percentage values in brackets state the fraction in relation to the number of authors of the row above. (1) College of Arts & Celtic Studies, (2) College of Business & Law, (3) College of Engineering, Mathematical & Physical Sciences, (4) College of Human Sciences, (5) College of Life Sciences. Data Source: UCD Research

For further analysis we switch from researchers to the analysis of publications. In Table 2 and Figure 2 we examine the level of co-authorship within and between the different colleges. Especially for the *College of Arts and Celtic Studies* and for the *College of Human Sciences* journal publications and proceedings are by far not the only research output. But it is still interesting to see that from the 61 researchers with output covered in the Web of Science there is no co-authorship within UCD. We can observe that 11% of co-authored publications from the *College of Human Sciences* are co-authored within UCD.

| College | All publications within UCD | Publications with one author from UCD | collaboration numbers (rates) | | | |
|---|---|---|---|---|---|---|
| | | | total | intra-school | inter-school | inter-colleges |
| Arts & Celtic Studies | 126 | 126 (100%) | 0 (0.0%) | 0 (0.0%) | 0 (0.0%) | 0 (0.0%) |
| Business & Law | 178 | 153 (86%) | 25 (14%) | 11 (6.2%) | 0 (0.0%) | 14 (7.9%) |
| Engineering, Mathematical & Physical Sciences | 2036 | 1782 (88%) | 254 (12%) | 186 (9.1%) | 6 (0.3%) | 62 (3.0%) |
| Human Sciences | 400 | 357 (89%) | 43 (11%) | 16 (4.0%) | 12 (3.0%) | 15 (3.8%) |
| Life Sciences | 5173 | 4248 (82%) | 925 (18%) | 593 (12%) | 269 (5.2%) | 63 (1.2%) |

Tab 2: Numbers of publications and collaboration rates per college between 1998-2007 within UCD, (1) College of Arts & Celtic Studies, (2) College of Business & Law, (3) College of Engineering, Mathematical & Physical Sciences, (4) College of Human Sciences, (5) College of Life Sciences.

We find the highest percentage of co-authorship within the *College of Life Sciences*, in which 18 % of the publications are co-authored. This college also shows the highest percentage of co-authorship within the schools 12 % and "small" interdisciplinarity between their schools with 5,2%. Whereas the *College of Engineering, Mathematical & Physical Sciences* have little co-authorship between the schools of this college, but with 3% a higher level of co-authorship with schools from other colleges, "big" interdisciplinarity. While, the total number of publications is much smaller for *the College of Business and Law* and it consists of only two schools, it is nonetheless interesting that the number of co-authorship between schools of different colleges is with 7.9%, "big" interdisciplinarity, even higher than the co-authorship within their own schools with 6.2 %. For the *College of Human Sciences* we find almost similar values for within-school co-authorship 4%, the co-authorship between schools of the same college 3%, "small" interdisciplinarity and between schools of different colleges 3.8%, "big" interdisciplinarity.

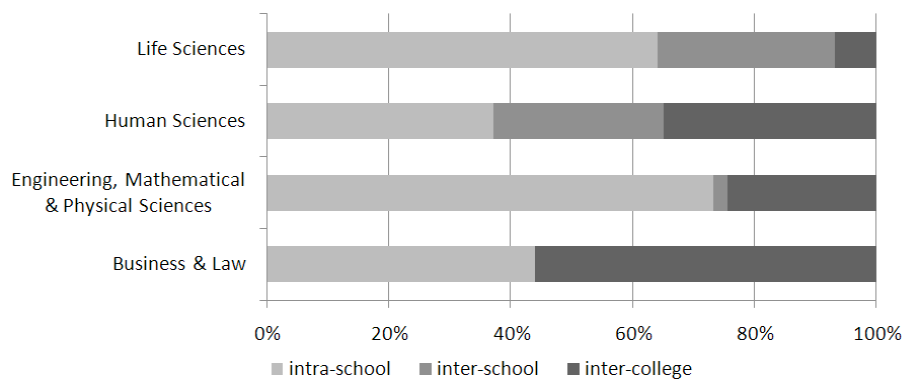

Fig 2: Collaboration rates in relation to the total collaboration rate for the four colleges with co-publications within UCD between 1998-2007.

**Methods II: Unit co-publication analysis**

To further investigate the collaboration on the school- and on the college-level we restrict our analysis to co-authored publications and focus on school/college affiliation of authors. Here the schools and colleges respectively replace the authors as vertices in the co-publication network. Since the analysis is identical for schools and colleges and is carried out for both separately, we call both 'units' $u_j$ in the course of this method description.

First the list of authors of each publication is transformed to a set of corresponding units (no multiples). Every unit-pairing within this set is taken as a co-publication link between these units. The set ensures that a unit-pairing is only taken into account once and if it contains only one element (in the case of a co-publication with authors affiliated to the same unit) a co-publication link from the unit to itself is established. The co-publication links are cumulated over all publications. Secondly we determine the share $s_{i\,j}$ of co-publications for each unit pairing $(u_i, u_j)$, where $(u_i, u_j)$ is equivalent to $(u_j, u_i)$. This creates a relative measure for intra- and inter-unit collaboration and is included as a percentage value in Table 3 & 4.

Additionally we compare these shares of co-publications with expected co-publication values stemming from a basic model of random-matching across units (Jones et al, 2008). This model considers the fraction of co-authored publication that includes unit $u_j$. More precisely for the random-matching process the fractions $f_j$ of pair-entries occupied by unit $u_j$ are the relevant input variables.

$$f_j = s_{jj} + \frac{1}{2} \sum_{i \neq j} s_{ij}$$

The expected probability $p_{ij}$ that a co-authored publication within one unit (i=j) or between two units includes units $u_i$ and $u_j$ results under random matching between the units to

$$p_{ij} = \begin{cases} f_i f_j & \text{if } i = j \\ 2 f_i f_j & \text{otherwise} \end{cases}$$

and can also be seen as the expected share of co-publications for the unit-pairing ($u_i$, $u_j$). A high expected share of co-publication between two units arises if their overall involvement in co-publication activity is large. We compare both shares taking the ratio of the actual share $s_{ij}$ to the expected share $p_{ij}$. These ratios are included in brackets in Table 3 & 4. A ratio > 1 stands for a higher disposition for co-publication than expected under random matching for this unit pairing. A ratio < 1 indicates a smaller likelihood for co-publication than expected between these units. For the interpretation of resulting expected shares one should keep in mind that the random matching model produces higher inter-unit than intra-unit co-publication shares. Notwithstanding it is appropriate to compare different share-ratios of inter-unit co-publication.

**Results II: Inter-school and inter-college co-publication analysis**

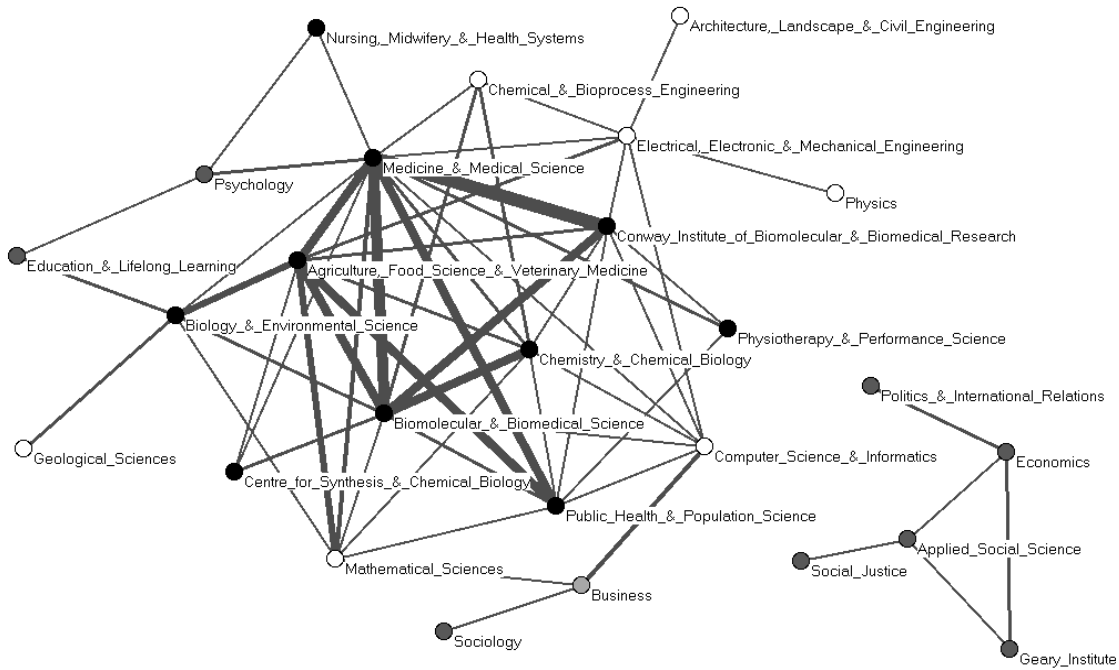

Fig 3: Co-authorship network of 26 UCD schools with co-publishing researchers between 1998-2007 (62% of all 42 schools). The levels of gray of nodes distinguish the colleges, from white to black: College of Engineering, Mathematical & Physical Sciences, College of Business and Law, College of Human Sciences, College of Life Sciences. The width of the lines is proportional to the radical of the number of co-authored publications between the corresponding schools, Software used: Pajek

In Figure 3 we focus on co-authorship between different schools to investigate interdiciplinarity. The thickness of the lines is proportional to the radical of the number of co-authored publications between the corresponding schools. The different shades of gray represent different colleges. We used Kamada-Kawai (with separated components), one of Pajek's 'spring embedded' network drawing algorithms, to enable meaningful representation of the school co-publication network.

The highest level of co-authorship between different schools, can be observed between the schools of the *College of Life Sciences*. The most central schools are *Medicine and Medical Sciences, School of Biomolecular & Biomedical Science, School of Agriculture, Food Science & Veterinary Medicine* and the *Conway Institute of Biomolecular and Biomedical Research*. From Figure 3 it is also possible to infer to which extent the classification of schools into colleges accords with the school-groups emerging from the collaboration activity. A high level of accordance is observable for the *College of Life Sciences* and for five schools of the *College of Human Sciences*. Schools of the *College of Engineering, Mathematical & Physical Sciences* seem to surround the *College of Life Sciences* and for the *School of Business* no other co-publishing school of the own college is available.

| Schools | Architecture, Landscape & Civil Eng. | Chemical & Bioprocess Eng. | Conway Institute of Biomolecular & Biomedical Research | Biology & Environmental Sc. | Agriculture, Food Sc. & Veterinary Med. | Med. & Medical Sc. | Geological Sc. | Business | Biomolecular & Biomedical Sc. | Computer Sc. & Informatics | Mathematical Sc. | Physics | Electrical, Electronic & Mechanical Eng. | Chemistry & Chemical Biology | Nursing, Midwifery & Health Systems | Public Health & Population Sc. |
|---|---|---|---|---|---|---|---|---|---|---|---|---|---|---|---|---|
| Architecture, Landscape & Civil Eng. | 0.97% (87.76) | 0.00% (0.00) | 0.00% (0.00) | 0.00% (0.00) | 0.00% (0.00) | 0.00% (0.00) | 0.00% (0.00) | 0.00% (0.00) | 0.00% (0.00) | 0.00% (0.00) | 0.00% (0.00) | 0.00% (0.00) | 0.16% (3.73) | 0.00% (0.00) | 0.00% (0.00) | 0.00% (0.00) |
| Chemical & Bioprocess Eng. | | 1.70% (36.96) | 0.00% (0.00) | 0.00% (0.00) | 0.00% (0.00) | 0.08% (0.08) | 0.00% (0.00) | 0.00% (0.00) | 0.32% (0.93) | 0.00% (0.00) | 0.00% (0.00) | 0.00% (0.00) | 0.08% (0.91) | 0.40% (3.24) | 0.00% (0.00) | 0.00% (0.00) |
| Conway Institute of Biomolecular & Biomedical Research | | | 0.49% (2.06) | 0.00% (0.00) | 0.40% (0.15) | 5.34% (2.19) | 0.00% (0.00) | 0.00% (0.00) | 2.43% (3.09) | 0.08% (0.14) | 0.00% (0.00) | 0.00% (0.00) | 0.08% (0.40) | 0.16% (0.57) | 0.00% (0.00) | 0.08% (0.30) |
| Biology & Environmental Sc. | | | | 2.35% (16.58) | 1.29% (0.61) | 0.16% (0.09) | 0.40% (4.58) | 0.00% (0.00) | 0.40% (0.66) | 0.00% (0.00) | 0.08% (0.47) | 0.00% (0.00) | 0.00% (0.00) | 0.00% (0.00) | 0.00% (0.00) | 0.00% (0.00) |
| Agriculture, Food Sc. & Veterinary Med. | | | | | 23.14% (2.94) | 1.94% (0.14) | 0.00% (0.00) | 0.00% (0.00) | 2.18% (0.48) | 0.00% (0.00) | 1.05% (0.81) | 0.00% (0.00) | 0.24% (0.21) | 0.40% (0.25) | 0.00% (0.00) | 2.18% (1.41) |
| Med. & Medical Sc. | | | | | | 17.48% (2.78) | 0.00% (0.00) | 0.00% (0.00) | 3.48% (0.86) | 0.08% (0.03) | 0.65% (0.56) | 0.00% (0.00) | 0.08% (0.08) | 0.49% (0.33) | 0.16% (0.28) | 1.86% (1.35) |
| Geological Sc. | | | | | | | 0.97% (70.54) | 0.00% (0.00) | 0.00% (0.00) | 0.00% (0.00) | 0.00% (0.00) | 0.00% (0.00) | 0.00% (0.00) | 0.00% (0.00) | 0.00% (0.00) | 0.00% (0.00) |
| Business | | | | | | | | 0.89% (41.96) | 0.00% (0.00) | 0.97% (5.42) | 0.08% (1.20) | 0.00% (0.00) | 0.00% (0.00) | 0.00% (0.00) | 0.00% (0.00) | 0.00% (0.00) |
| Biomolecular & Biomedical Sc. | | | | | | | | | 2.18% (3.34) | 0.08% (0.08) | 0.08% (0.22) | 0.00% (0.00) | 0.00% (0.00) | 2.35% (4.98) | 0.00% (0.00) | 0.24% (0.55) |
| Computer Sc. & Informatics | | | | | | | | | | 5.34% (14.12) | 0.00% (0.00) | 0.00% (0.00) | 0.08% (0.32) | 0.16% (0.45) | 0.00% (0.00) | 0.16% (0.48) |
| Mathematical Sc. | | | | | | | | | | | 1.21% (22.83) | 0.00% (0.00) | 0.00% (0.00) | 0.16% (1.20) | 0.00% (0.00) | 0.08% (0.64) |
| Physics | | | | | | | | | | | | 3.24% (29.41) | 0.16% (1.18) | 0.00% (0.00) | 0.00% (0.00) | 0.00% (0.00) |
| Electrical, Electronic & Mechanical Eng. | | | | | | | | | | | | | 1.62% (38.02) | 0.00% (0.00) | 0.00% (0.00) | 0.00% (0.00) |
| Chemistry & Chemical Biology | | | | | | | | | | | | | | 0.81% (9.54) | 0.00% (0.00) | 0.08% (0.50) |
| Nursing, Midwifery & Health Systems | | | | | | | | | | | | | | | 0.97% (75.67) | 0.00% (0.00) |
| Public Health & Population Sc. | | | | | | | | | | | | | | | | 0.32% (4.28) |

Tab 3: Shares of co-publications between schools (percentage values) and their ratio to shares that are expected according to a random matching model (values in brackets). The matrix was calculated for all 42 schools of UCD but included in the table are 16 schools with an overall fraction of co-publications > 1%. Colored are cells with a high co-publication share:
Small interdisciplinary (same college), higher than expected, lower than expected,
Big interdisciplinary (different college), higher than expected, lower than expected,

Within Table 3 we analyze the relative co-publication output between two schools and compare the actual collaboration shares to shares that are expected according to a random matching model. Additionally we look on differences between "small" and "big" interdisciplinary inter-school links.

The collaboration seems to be small between the *School of Business* and *Schools of Computer Science & Informatics* but the actual collaboration share is 5.4 times higher than expected. Also between the *School of Chemistry & Chemical Biology* and the *School of Biomolecular & Biomedical Science* the actual collaboration share is 4.98 times higher than expected. On the other hand we found links which appear very strong within Fig. 3 and have a high share of co-publications for example the *School of Agriculture, Food Science & Veterinary Medicine* and the *School of Medicine and Medical Science* although they have a much lower (0.14) co-publication share than expected.

Beside these findings some more general conclusions can be drawn analyzing Table 3:
- Generally intra-school cooperation shares are higher than expected. This is due to the high intra-school cooperation rates of the big colleges *Life Science* and *Engineering, Mathematical & Physical Sciences* (cf. Fig. 2) and due to the tendency of the random matching model to produce smaller intra-school cooperation shares.
- High actual co-publication shares do not directly imply a relatively high likelihood for co-publication between these schools. For example within the college of *Life Science* some of the high co-publication shares (cf. Fig. 3) are rather small in relation to the expected values (e.g. *School of Agriculture, Food Science & Veterinary Medicine* with the *School of Medicine and Medical Science* as well as with the *School of Biomolecular & Biomedical Sciences*).
- Even though the actual inter-school co-publication shares are on average larger for schools of the same college than for inter-college cooperation (2.56% vs. 0.63%)[1], their averaged ratio to expected shares does not show this difference (1.68 vs. 2.59). This means that for the investigated intra-college and inter-college school-pairings no general difference in tendency for collaboration is observable between "small" and "big" interdisciplinary links. The difference in actual shares is mainly due to a difference in overall co-publication activity.

| Colleges | Arts & Celtic Studies | Engineering, Math. & Phys. Sc. | Business & Law | Life Sc. | Human Sc. |
|---|---|---|---|---|---|
| Arts & Celtic Studies | 0.00% (1.00) | 0.00% (1.00) | 0.00% (1.00) | 0.00% (1.00) | 0.00% (1.00) |
| Engineering, Math. & Phys. Sc. | | 16.41% (4.52) | 1.11% (1.89) | 4.19% (0.14) | 0.00% (0.00) |
| Business & Law | | | 0.94% (39.72) | 0.00% (0.00) | 0.09% (0.92) |
| Life Sciences | | | | 73.68% (1.26) | 1.20% (0.26) |
| Human Sciences | | | | | 2.39% (25.99) |

Tab 4: Shares of co-publications between college (percentage values) and their ratio to shares that are expected according to a random matching model (values in brackets).

Analogical to the schools cooperation we analyze the co-publication shares between colleges (Table 4). We observe a higher share, i.e., 1.89 than expected between the *College of Business and Law* and *College of Engineering, Mathematical & Physical Sciences* which has its counterpart in the collaboration of schools belonging to these colleges (cf. Tab 3 *School of Business* and *School of Computer Science & Informatics*). Between the *College of Life Sciences* and *College Engineering, Mathematical & Physical Sciences* we have a lower rate than expected, 0.14, however these two colleges have a high co-publication share of the total. Again the inter-college cooperation shares are higher than expected (except for the *College of Arts & Celtic Studies* with no co-publications) which is again due to the high intra-college collaboration rates particularly of the big colleges (cf. Fig. 2).

---

[1] The values were averaged over the marked cells of Table 3 for small and big interdisciplinary links respectively. We restricted our analysis to intra-college links with an actual co-publication share > 1% and to inter-college links with a share > 0.3%.

**Results III: Evolution of co-publication activity within UCD**

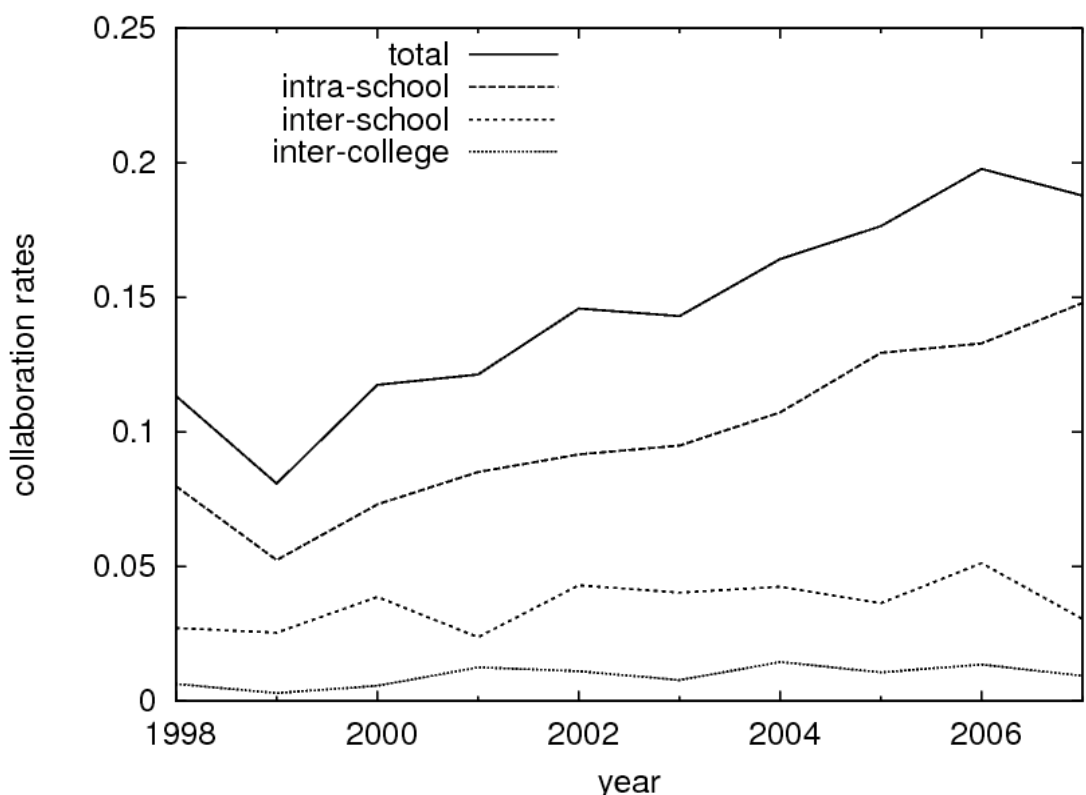

Fig 4: Evolution of co-publications within UCD

From the 7836 papers analyzed between 1998 and 2007, 1170 were co-published within University College Dublin. Within the same schools were about 806 publications co-published and between different schools but within same college 287 publications (cf. Table 2). Between different schools and between different colleges we found 77 co-authored publications.
In 1998 the collaboration rate was 11%, constantly increasing to highest level 19.8% in 2006. So the collaborative structure within the university almost doubled. However, the overall observed growth is mainly limited to the intra-school co-authorship.

**4. Discussion**

Our data shows that the rate of co-authorship at UCD has almost doubled within a decade. Growing from 11% co-authorship in 1998 to 19.8 % 2006. However this is mainly caused, by the increasing level the of intra-school collaboration, whereas small interdisciplinarity, collaboration of schools within one colllege, as well as big interdisciplinarity, collaboration between schools of different colleges stays almost at the same level since 2001.

Using co-authorship as an indicator for mapping the interdisciplinary cooperation structure of the university shows that within and between colleges and schools the co-publication activity differs.
We observe for example:

- The highest amount of intra-school and inter-school co-authorship we can observe within the *College of Life Sciences*. Compared to that the *College of Engineering, Mathematical & Physical Sciences* has a lower level of intra school co-authorship, a very low level between the schools, but interestingly a higher level of inter-colleges co-authorship shares. Whereas some of the schools of the *College of Life Sciences* form a cluster, the *schools of Engineering, Mathematical & Physical Sciences* are connected around the schools (big-interdisciplinarity, cf. Fig 3).
- We compared within the schools of the *College of Life Sciences* the shares of co-publications with expected co-publication values stemming from a basic model of random-matching across units. We got higher values as expected as well as lower providing more insight than the actual shares. The collaboration between the *School of Biomolecular & Biomedical Science* and the *School of Chemistry and Chemical Biology* is 4.98 times higher than expected.

- For collaborations between schools from different colleges (big interdisciplinarity): *School of Computer Science and Informatics* and *School of Business* the collaboration is 5.42 times higher than expected. Between the *School of Geological Science* and the *School of Biology and Environmental Sciences* it is 4.58 times higher than expected. The collaboration between the *School of Chemistry and Chemical Biology* and the *School of Chemical and Bioprocess Engineering* it is 3.24 times higher than expected (even the similarity of the schools names' suggest that these schools are not that thematically far away as the college structure might suggest).

These observations lead us to assume that there are specific profiles of schools and individuals with skills crucial to connect and contribute to other schools/individuals. This ranges from cooperation between thematically close areas to bridging huge disciplinary distances. We can easily identify "champions" of small and big interdisciplinary co-publication, brokers between disciplines, and areas where this is not working at all. We also found some evidence for the findings from Cummings and Kiesler, 2008 that big interdisciplinary collaboration is often performed and maintained from few individuals who already paid the start-up costs of getting to know each other's languages, approaches and methodologies.

However, the kind of conclusions we can draw from these results need to be put into context. The *College of Business and Law* consisted only of two schools which interferes the level of observable co-authorship between schools in this college. Further to this, our approach did not take into account the educational backgrounds of the individuals related to the schools. If we assume that (some) schools try to foster interdisciplinary research by incorporating researchers from different backgrounds, this trend would not be visible within our approach. (The authors of this paper are related to the Business School but their background is sociology and physics).

We also have to take into account that Scientific disciplines are differently represented in databases such as the Web of Science (WoS). This biases our data especially, the results based on absolute publication numbers. To interpret these results appropriately, we need to relate them to the different publication cultures in the scientific disciplines.

These differences manifest themselves in various aspects, e.g. (co-)authorship is less frequent in theoretical than in experimental fields etc. In this context, we can only illustrate this point using some anecdotal evidence for different disciplinary publication cultures: van Raan (2005) estimates that in the medical sciences the share of publications covered by the Citation Index is about 80 to 95 percent, whereas in the social sciences it is much lower, in psychology 50-70 per cent coverage, in the engineering fields it is about 50 per cent.

We stated absolute and relative values and analyzed mainly relative ones which are less biased. However you could argue that in different disciplines the collaboration has also different forms, e.g. shared workshops, conferences, exhibitions. Furthermore, Laudel (2002) also shows that co-publication as an indicator does not cover important aspects of research collaboration: most collaboration within a university is never rewarded by a co-authored publication. This means that, even if the data is weighted against scientific publication cultures, co-publication would need to be complemented by other indicators of interdisciplinary cooperation dynamics. Laudel and Gläser (2006) have presented how and where scientific practice of research collaborations slips through the nets of current evaluation instruments. This particularly refers to using publications as the only measure for assessing research collaboration and performance within an over-simplified and incorrectly standardised evaluation approach. On the other hand it seems reasonable und sustainable to base individual and organizational development on measurable outputs like publications. We hope we were able to demonstrate that these kind of studies provide a useful picture for university development.

**Acknowledgement**

This paper is based on research sponsored by UCD Research. The authors would like to thank the reviewers for their comments and Michael J. Barber and Conrad Lee for a helpful discussion.